\documentclass[conference]{IEEEtran}
\IEEEoverridecommandlockouts
\usepackage{cite}
\usepackage{amsmath,amssymb,amsfonts}
\usepackage{algorithmic}
\usepackage{graphicx}
\usepackage{textcomp}
\usepackage{xcolor}
\usepackage{amsmath}
\usepackage[lined,boxed,ruled]{algorithm2e}
\usepackage{soul}
\usepackage{comment}
\usepackage{subcaption}

\def\BibTeX{{\rm B\kern-.05em{\sc i\kern-.025em b}\kern-.08em
    T\kern-.1667em\lower.7ex\hbox{E}\kern-.125emX}}

\definecolor{Red}{rgb}{0,0,0}
\definecolor{Blue}{rgb}{0,0,0}
\definecolor{Green}{rgb}{0,0,0}
\definecolor{Gray}{gray}{0}

\newcommand{\blue}[1]{\textcolor{Blue}{#1}}

\begin{document}

\title{Learning a Distributed Control Scheme for Demand Flexibility in Thermostatically Controlled Loads
\thanks{\IEEEauthorrefmark{1}This material is based upon work supported by the U.S. Department of Energy’s Office of Energy Efficiency and Renewable Energy (EERE) under the Building Technologies Office Award Number DE-EE0007682} 
}
\author{\IEEEauthorblockN{Bingqing Chen\IEEEauthorrefmark{4},
Weiran Yao\IEEEauthorrefmark{4}, 
Jonathan Francis\IEEEauthorrefmark{2}\IEEEauthorrefmark{3} 
and Mario Berg\'{e}s\IEEEauthorrefmark{4}}
\IEEEauthorblockA{\IEEEauthorrefmark{4} Department of Civil and Environmental Engineering,
Carnegie Mellon University, Pittsburgh, PA 15213, USA}
\IEEEauthorblockA{\IEEEauthorrefmark{2} School of Computer Science,
Carnegie Mellon University, Pittsburgh, PA 15213, USA}
\IEEEauthorblockA{\IEEEauthorrefmark{3} Bosch Research \& Technology Center, Pittsburgh, PA 15222, USA\\
\texttt{\{bingqinc, wyao1, jmf1, mberges\}@andrew.cmu.edu}}}
\IEEEoverridecommandlockouts
\IEEEpubid{\makebox[\columnwidth]{
 978-1-7281-6127-3/20/\$31.00~\copyright 2020 IEEE
\hfill} \hspace{\columnsep}\makebox[\columnwidth]{ }}
\maketitle
\IEEEpubidadjcol

\begin{abstract}
Demand flexibility is increasingly important for power grids, in light of growing penetration of renewable generation. Careful coordination of thermostatically controlled loads (TCLs) can potentially modulate energy demand, decrease operating costs, and increase grid resiliency. However, it is challenging to control a heterogeneous population of TCLs: the control problem has a large state action space; each TCL has unique and complex dynamics; and multiple system-level objectives need to be optimized simultaneously. To address these challenges, we propose a distributed control solution, which consists of a central load aggregator that optimizes system-level objectives and building-level controllers that track the load profiles planned by the aggregator. To optimize our agents' policies, we draw inspirations from both reinforcement learning (RL) and model predictive control. Specifically, the aggregator is updated with an evolutionary strategy, which was recently demonstrated to be a competitive and scalable alternative to more sophisticated RL algorithms and enables policy updates independent of the building-level controllers. We evaluate our proposed approach across four climate zones in four nine-building clusters, using the newly-introduced \texttt{CityLearn} simulation environment. Our approach achieved an average reduction of 16.8\% in the environment cost compared to the benchmark rule-based controller.
\end{abstract}




\section{Introduction}
\label{sec:introduction}

Whereas renewable energy resources present enormous opportunities for reducing the grid's reliance on fossil fuels, they also presents \textit{new} challenges for grid operators to balance supply and demand, due to their intermittent and variable nature.  For instance, in areas with high solar adoption, generators need to quickly ramp up when the sun sets \cite{duckcurve2017}. 



Traditionally, the load from the demand side is viewed as a given and the supply side manages the power generation to match it \cite{neukomm2019grid}. 
However, this paradigm is no longer cost-effective. Demand side resources can provide flexibility to the grid by reducing or shifting their loads in response to price or direct control signals \cite{neukomm2019grid}.  
Specifically, residential thermostatically controlled loads (TCLs), such as air conditioners, refrigerators, and electric water heaters account for 20\% of all electricity consumption in the United States \cite{hao2014aggregate}, and due to their inherent flexibility from thermal inertia, they can provide various grid services without compromising their end uses. 

Despite the potential of TCLs for grid services, there are several challenges to utilizing this potential. Firstly, for TCLs to be a meaningful resource to the grid, their inherent flexibility must be aggregated over a population \cite{neukomm2019grid}; this yields a control problem with a large state action space. A common solution is centralized control of an aggregate model \cite{burger2017generation}, but we discuss its limitations in Section \ref{sec:IIA}. Secondly, the TCL population are generally heterogeneous in sizes and configurations. At the same time, each TCL has complex dynamics, device-specific constraints, and is subject to stochastic usage patterns \cite{kazmi2016generalizable}. %
Finally, many grid objectives may need to be optimized simultaneously|some of which are competing, e.g., efficiency vs. flexibility \cite{maasoumy2014model}. Other objectives may need to be optimized over a long time horizon (e.g., monthly peak demand \cite{wang2015near}) or do not permit analytical solutions. We elaborate on these challenges and summarize related work in Section \ref{sec:review}.

\begin{figure*}
  \includegraphics[width=\textwidth]{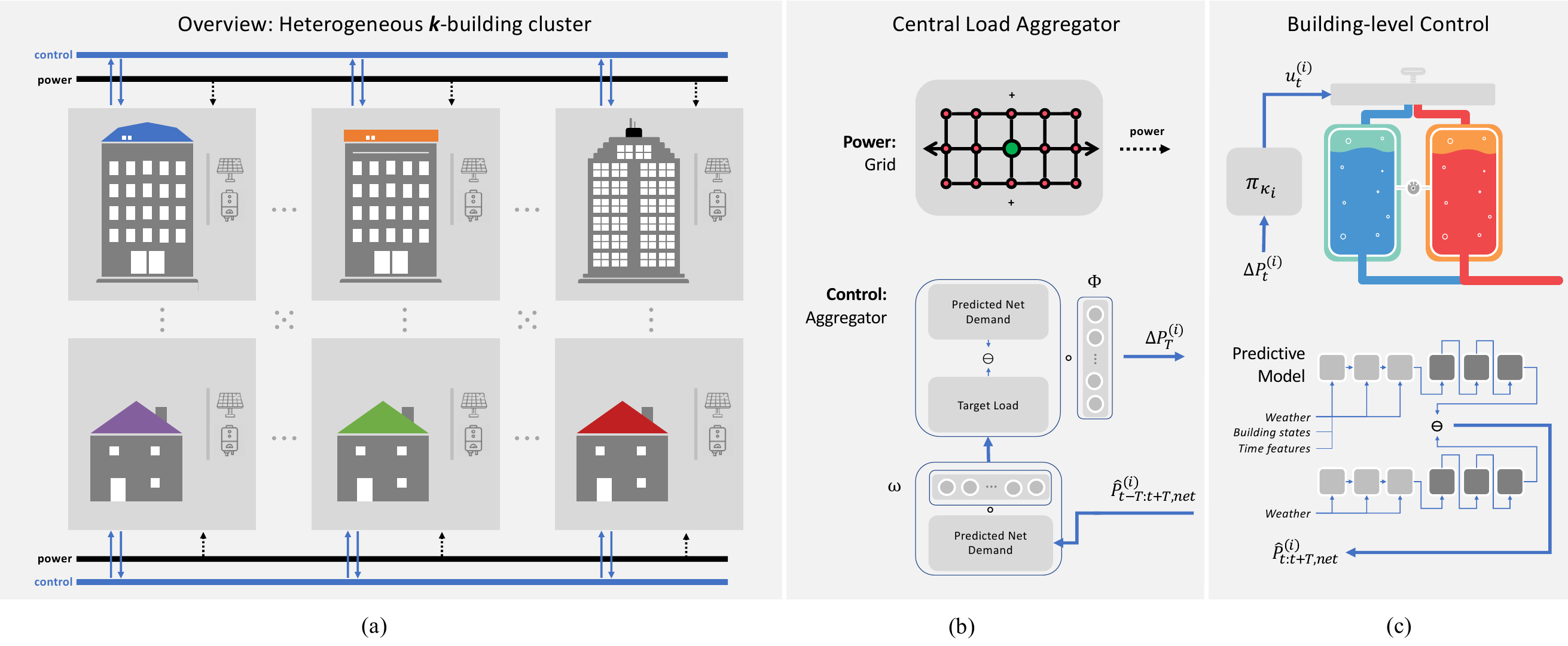}
  \caption{System overview. A heterogeneous cluster of $k$ buildings in (a) is connected to the power grid and managed by the load aggregator in (b). Each building has controllable TCLs, illustrated in (c, top). The building $i$ predicts its net energy demand over a planning horizon, i.e., $\hat{P}^{(i)}_{t:t+T, net}$, using its predictive model (c, bottom). The aggregator collects the predictions from the building cluster and plans for a target load based on a learnable filter $\omega$. The difference between the target load and aggregated net load is apportioned to each building, with a learnable vector $\Phi$, the result of which is the control command $\Delta P^{(i)}$ to each building. Each building matches the control command based on its policy $\pi_{\kappa_i}$. 
   \label{fig:framework}}
\end{figure*}

To alleviate these challenges, we present a learning-based, distributed solution for controlling a heterogeneous population of TCLs to provide grid services.  
Instead of directly optimizing the task objectives over the entire system, we break down the problem into more tractable sub-problems. Our framework consists of a central load aggregator and building-level controllers for each building. The load aggregator plans for a load profile that is desirable for the grid and apportions it to each building, thereby simplifying the objective of each building-level controller to that of a reference-tracking problem. To optimize the agents' policies, we draw inspirations from both the reinforcement learning (RL) and model predictive control (MPC) literature. Since the system-level objectives may be difficult to optimize analytically, we find an approximate solution for the \textit{aggregator} with RL. Such approach is generalizable to different grid objectives. Specifically, we use a gradient-free RL algorithm from the class of nature-inspired evolutionary strategies (ES). This allows us to update the aggregator, independent of the \textit{building-level controllers}. To improve sample efficiency \cite{chen2019gnu}, we utilize domain knowledge and model each TCL as a virtual battery \cite{hao2014aggregate, zhao2017geometric}. Thus, the reference-tracking problem can be solved efficiently with a quadratic program (QP). At the same time, we account for heterogeneity and complexity in system dynamics by adaptively learning model parameters of each TCL with prediction error minimization (PEM), a common approach for system identification \cite{ljung2001system}.

We evaluate our approach using the newly-introduced \texttt{CityLearn} environment \cite{vazquez2019citylearn}, where the task is to control thermal storage units in a heterogeneous building cluster. The environment's objective is defined as the average of net electricity consumption, 1-load factor, ramping, average daily peak demand, and annual peak demand|normalized by those of a rule-based controller (RBC). We use four nine-building clusters, located in four anonymized climate zones, and achieve a 16.8\% average reduction in the environment cost, compared to the benchmark RBC. We also compare our approach to model-free RL baselines and demonstrate the benefit of incorporating prior knowledge of system dynamics. 


\section{Related Work}
\label{sec:review}

\subsection{Architectures for TCL control} \label{sec:IIA}
The primary challenge for jointly controlling a large number of TCLs is the large state action space. To address this challenge, a popular approach in the model-based control literature is to develop an aggregate model for the population and control the population in a centralized manner. Examples of such aggregate model include the state bin transition model in \cite{koch2011modeling, zhang2013aggregated} and the virtual battery model in \cite{hao2014aggregate, zhao2017geometric}. However, these aggregate models depend on the assumptions that each system may be characterized by 1\textsuperscript{st} (or 2\textsuperscript{nd} \cite{zhang2013aggregated}) order linear model, and that all systems in the population share the same model structure and control scheme. These aggregate models have low fidelity and do not capture system specific dynamics \cite{burger2017generation}. Specifically, it was demonstrated in \cite{xu2014modeling} that 1\textsuperscript{\textit{st}} and 2\textsuperscript{\textit{nd}}-order models failed to accurately capture the thermodynamics of an individual electric water heater. Aside from the centralized architecture, decentralized control \cite{tindemans2015decentralized} and distributed control \cite{burger2017generation} approaches have also been proposed in the literature.
The key advantage of a decentralized control approach is that each system can be controlled based on local information, i.e. no communication is necessary. However, the applications of decentralized control methods are thus limited to frequency regulation and real-time load shaping \cite{burger2017generation}. In a distributed architecture, which we also utilize, each system in responsible for its own control, and coordinates with others to achieve a grid-level objective. 

\subsection{Reinforcement Learning for TCL control}
Given the difficulty in developing high-fidelity model for each system, RL has also been applied to controlling TCLs in works such as \cite{kazmi2016generalizable, ruelens2016reinforcement, claessens2018model, liu2006theory}. It is worth-noting that \cite{kazmi2016generalizable} and \cite{ruelens2016reinforcement} validated their approaches on individual electric water heaters in real-world settings. However, the sample complexity increases with the state action space \cite{liu2006theory}, and thus it may take an impractical amount of training time for grid-scale application without incorporating domain knowledge.

 Similar to this work, \cite{wang2015near, claessens2018model} combine RL and model-based control to improve the sample efficiency. To address the challenge of optimizing the monthly peak demand, i.e. the long planning horizon, \cite{wang2015near} proposed a near-optimal solution, where the charging / discharge of an energy storage unit was determined analytically by a model-based controller over each day, and the residual energy at the end of each day was approximated by Q-learning. To account for the large state action space,  \cite{claessens2018model} used Q-learning to find the aggregate action for the TCL population and then dispatched the aggregate action to individual units with proportional–integral control.

\subsection{Optimization Objectives for Distributed TCL control}

A variety of objectives have been discussed in the literature, such as: cost minimization \cite{liu2006theory}, energy efficiency \cite{kazmi2016generalizable}, day-ahead scheduling \cite{ruelens2016residential}, demand response \cite{zhang2013aggregated, tindemans2015decentralized}, frequency regulation \cite{hao2014aggregate, zhao2017geometric}, and peak demand reduction \cite{wang2015near}. However, the  approaches in these works were generally formulated based on their specific use case and may not generalize to alternative objectives. Furthermore, few work simultaneously optimize over more than two objectives.

\section{Approach} \label{sec:approach}

Our distributed control framework (Figure \ref{fig:framework}) consists of a load aggregator and building-level controllers. In Section \ref{sec:aggregator}, we discuss how the aggregator plans for the load profile to optimize grid-level objectives and updates its policy with an ES. In Section \ref{sec:pred_model}, we describe the predictive model for net energy demand, which is a component of the building-level controller. In Section \ref{sec:controller}, we describe the MPC strategy used by building-level controllers. 

\subsection{Central Load Aggregator}\label{sec:aggregator}
To optimize system-level objectives, we apply a learnable convolutional filter, $w\in \mathbb{R}^{2T+1}$, on the aggregate energy demand from $t-T$ to $t+T$ to get a target load,  $\tilde{P}_t$ (Eq.~\ref{eq:e_tilde}). $\hat{P}_{t, net}^{(i)}$ denotes the predicted net energy demand by building i at time t, assuming the TCLs only maintain their temperature at setpoint. T is the planning horizon, and $\mathcal{I}$ denotes the set of building indices. In this work, we use a planning horizon of 12 hours, i.e. T=12, and re-plan at each time-step based on new observations from the environment. The load that needs to be shifted, $\Delta P_t$, is the difference between the target load and the aggregate energy demand (Eq. \ref{eq:delta_e}). Similar to \cite{mahdavi2017model}, we apportion  $\Delta P_t$ over the cluster with a learnable weight vector $\Phi$ (Eq. \ref{eq:apportion}), where $\sum_i \Phi_i=1$. $\Phi$ corresponds to the relative percentage of flexibility a building has in relation to the building cluster. 
\begin{subequations}

\begin{equation}
    \label{eq:e_tilde}
    \tilde{P}_t = \sum_{l = -T}^{T} \sum_{i\in\mathcal{I}} \omega_l \hat{P}^{(i)}_{t+l, net}
\end{equation}

\begin{equation}
    \label{eq:delta_e}
    \Delta P_t = \tilde{P}_t-\sum_{i\in\mathcal{I}}\hat{P}^{(i)}_{t, net}
\end{equation}

\begin{equation}
    \label{eq:apportion}
    \Delta P_t^{(i)} = \Phi_i \Delta P_t
\end{equation}
\end{subequations}

One challenge in updating the policy of the load aggregator is that it depends not only on its own parameter, but also on that of each building's local controller's. Thus, a gradient-free algorithm, such as ES, is well suited to optimize $\Phi$ and $\omega$ independent of the building-level controllers. ES are black-box optimization algorithms inspired by natural evolution. Recent work has demonstrated ES to be a scalable \cite{salimans2017evolution} and competitive \cite{mania2018simple} alternative to other more sophisticated RL methods, rekindling research interest in ES. Some well-known ES approaches include Cross-entropy Method (CEM) \cite{szita2006learning},  
Natural Evolutionary Strategies (NES) \cite{salimans2017evolution}, and Finite Difference method \cite{mania2018simple}. The objective of ES is to find policy parameter $\theta$ that maximizes expected cumulative reward, $F(\theta)$. Unlike policy gradient methods, it is not necessary to take derivatives through the policy,  exemplified by the update rule of NES (Eq. \ref{eq:gradient}).
\begin{equation}
\nabla_\theta \mathbb{E}_{\theta\sim N(\mu, \sigma^2 I)}F(\theta) =  \frac{1}{\sigma}\mathbb{E}_{\epsilon\sim N(0, I)}F(\theta+\sigma\epsilon )\epsilon
\label{eq:gradient}
\end{equation}
Our approach is primarily based on NES \cite{salimans2017evolution}. we also incorporate a modification proposed in \cite{mania2018simple}, i.e., we adaptively select the update step size by normalizing with the standard deviation of the rewards collected in N rollouts, $\sigma_R$, instead of the exploration noise, $\sigma$. We initialize $\omega$ as a moving average smoother and $\Phi$ assuming that flexibility is proportional to the aggregate energy demand of a building. We summarize the control strategy of the aggregator and the update of its policy in Algorithm \ref{alg:es}, where the policy parameter $\theta=\{\Phi, \Omega\}$. The hyperparameters are $\alpha = 0.01$, $\sigma = 0.01$, $N = 4$.

\begin{algorithm}[ht]
\SetAlgoLined
\SetKwFunction{PredictNetConsumption}{predictConsumption}
\textbf{Input: } Learning rate $\alpha$, noise standard deviation $\sigma$, number of rollouts $N$, initial policy parameters $\theta_0$, policy of building i, $\pi_{\kappa_i}$\\
\textbf{Initialization: } Current policy parameters $\theta = \theta_0$\\
 \For{d = 0, $\dots$, \# Episodes (Days)} 
 {$\epsilon_d \sim \mathcal{N}(0, 1)$, $\mathbf{\theta}_d = \theta + \sigma \epsilon_d$\\
  \For{t = 0, $\dots$, 23, \# Steps (Hours)} 
  {$\hat{P}^{(i)}_{t:t+T, net}$ = \PredictNetConsumption($\mathbf{x}_t$);\\
   $\tilde{P}_t = \sum_{l = -T}^{T} \sum_{i\in\mathcal{I}} \omega_l \hat{P}^{(i)}_{t+l, net}$
$\Delta P_t = \tilde{P}_t-\sum_{i\in\mathcal{I}}\hat{P}^{(i)}_{t, net}$;
$\Delta P^{(i)}_t = \Phi_i \Delta P_t$

  \For{i = 0, $\dots$, \# Buildings} 
  {$u^{(i)}_{t} = \pi_{\kappa_i}(\Delta P^{(i)}_t)$\\}
  $x_{t+1}, r_{t+1} = \text{env.step}(u_t)$\\
  }
  Compute episodic return ${F}_d$\\
  Every $N$ episodes (days) update $\theta$:\\
  \hspace{0.5cm}  
  {$\theta \xleftarrow{} \theta + \alpha \frac{1}{p\sigma_R} \sum_{d\in\mathcal{D}}{F}_d$}
 }
 \caption{Load Aggregator with NES (Modified from \cite{salimans2017evolution, mania2018simple})}
 \label{alg:es}
\end{algorithm}

\subsection{Predictive Modeling} \label{sec:pred_model}


Each building has a predictive model for its net energy consumption over a planning horizon, i.e., $\hat{P}^{(i)}_{t:t+T, net}$. We assume that historical data are available to pre-train the predictive models. We use sequence-to-sequence (Seq2Seq) models for prediction. Seq2seq models consist of encoders that embed source sequences into hidden vectors and turn them into target sequences with a decoder model \cite{yao2020learning}. Bilinear attention mechanisms proposed in \cite{bahdanau2014neural} are employed in the decoder to select the input sequence dynamically.

We decompose the prediction task into two sub-task models: electric load predictor and renewable (specifically solar) generation predictor. The intuition for the decomposed design is that solar generation per unit is determined by weather conditions only (e.g., solar radiation, temperature, etc.), while electricity demand are impacted by other variables such as building attributes, past building states and resident's behaviors, etc. Finally, net electricity consumption can be computed by combining the outputs from two models in Eq.~\ref{eq:net_cal}.

\begin{equation}\label{eq:net_cal}
    P_{t,\text{net}}^{(i)} = P_{t, \text{total}}^{(i)} - C^{(i)}_{sol} P_{t, \text{gen}} 
\end{equation}
where $P_{t,\text{net}}^{(i)}$ is the net electricity consumption of building $i$ from the grid at time $t$, $P_{t, \text{total}}^{(i)}$ is the total electricity demand, $C^{(i)}_{sol}$ is the solar power capacity installed (kW) at building i and $P_{t, \text{gen}}$ is solar generation per unit. 



\subsubsection{Electric Load Predictor}

\begin{figure}[h]
   \begin{subfigure}[]{\linewidth}
     \centering
    \includegraphics[width = \linewidth]{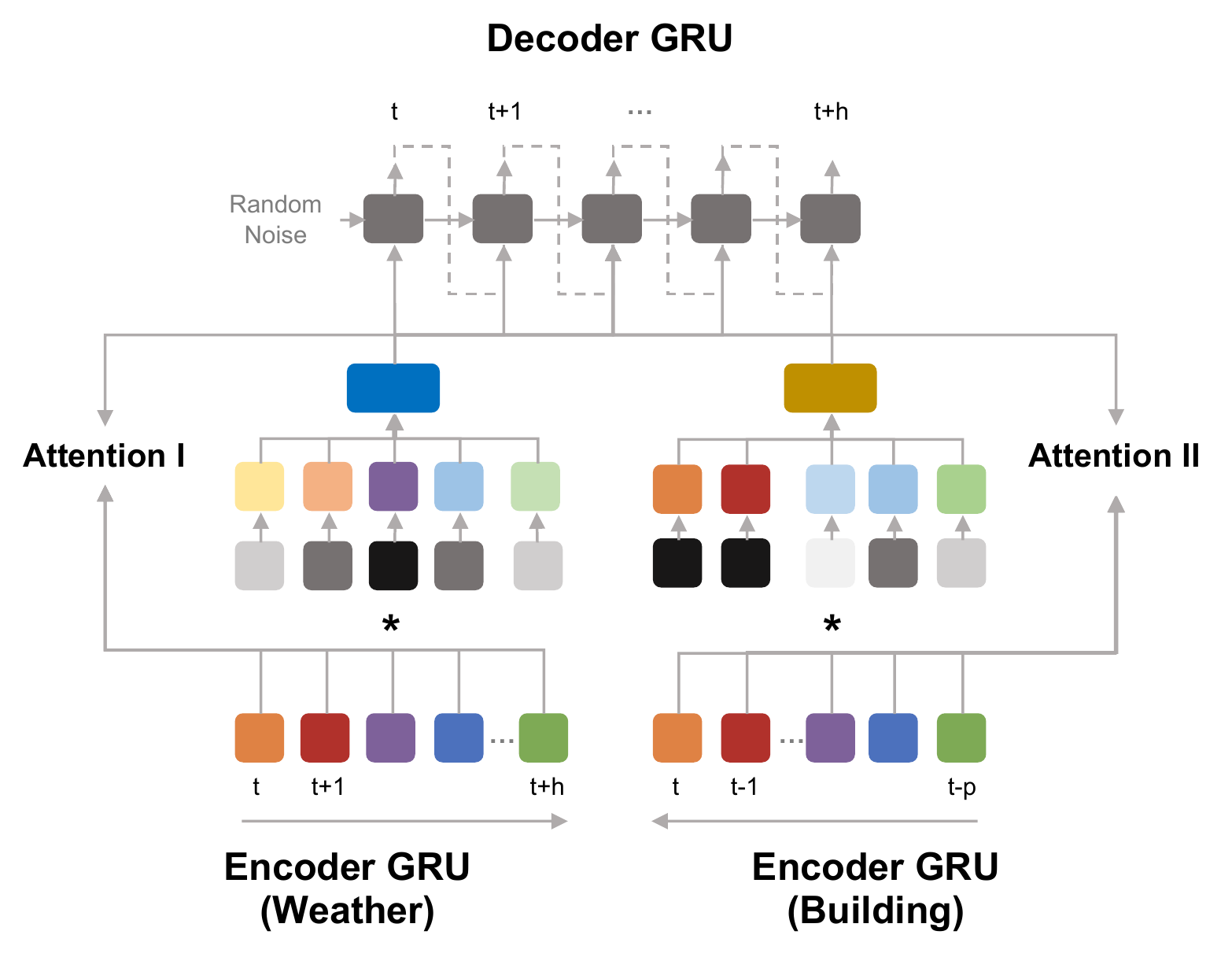}
    \caption{Electric load predictor without thermal storage.}
    \label{fig:gru-net}
    \end{subfigure}
    
  \begin{subfigure}[]{\linewidth}
    \centering
    \includegraphics[width = 0.65\linewidth]{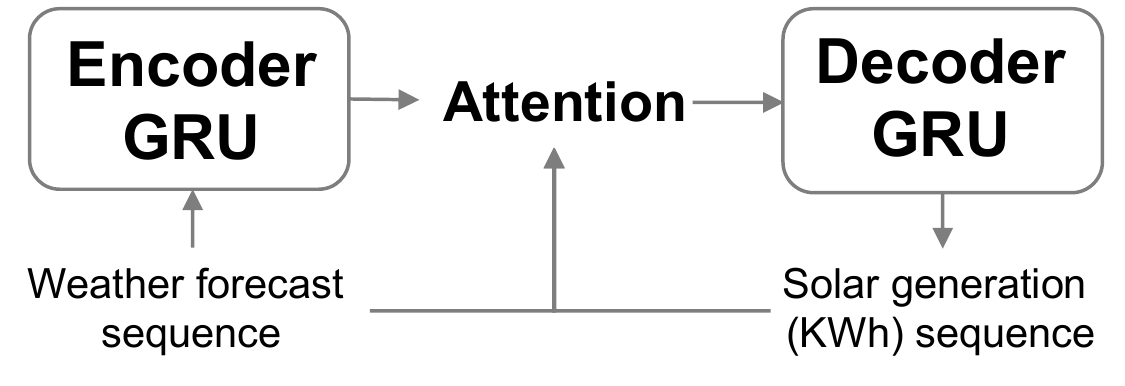}
    \caption{Solar generation predictor.}
    \label{fig:gru-solar}
    \end{subfigure}
\caption{Neural architecture for predictive models.}
\end{figure}

The electric load predictor triggers predictions of building total load 12 hours ahead. 
As shown in Figure \ref{fig:gru-net}, both encoders and decoder of the model use Gated Recurrent Unit (GRU \cite{cho2014learning}) as recurrent layers. The encoder includes a weather encoder for weather sequences, and a building encoder for processing lagged building states. We include static building attributes as part of building state $\mathbf{x}^b_t$ inputs at each time step. Time features are appended to both weather and building state inputs $[\mathbf{x}^w_t, \mathbf{x}^b_t]$ to encode time-dependent information of every building and weather state. The decoder employs two independent attention models to extract and attend to hidden states of weather and building encoders. 
The output of the model at each time step 
is then used as inputs of the next time step autoregressively.

\subsubsection{Solar Generation Predictor}
We use Seq2Seq neural model for translating the interpolated weather forecast into solar generation $P_{t, \text{gen}}^{(i)}$. As shown in Figure \ref{fig:gru-solar}, the encoder and decoder are both GRUs. Similarly, time features are appended to weather inputs $\mathbf{x}^w_t$ to embed time-dependent information. Bilinear attention models are employed to attend to weather forecast sequences for predicting solar generation.

\subsubsection{Hyperparameters and Training} For GRUs in the electric load predictor, we use: Tanh
activation, 128 hidden dimensions, 1 layer and recurrent dropout of 0.75. For GRUs in solar generation predictor, we use: Tanh activation, 32 hidden dimensions, 1 layer and recurrent dropout of 0.5. Attention has 128 hidden states. We train the network using Adam \cite{kingma2014adam} to optimize mean-squared-error (MSE) for a maximum of 50 epochs and early-stops if validation error does not decrease for 2 epochs. Learning rate of 0.001, teacher-forcing ratio of 0.5 and mini-batch size of 64 are used.  

\subsection{Building-level Controller}\label{sec:controller}
Each building is operated by a local controller that tracks the command, \blue{$\Delta P^{(i)}$} from the load aggregator. By modeling each TCL as a \blue{virtual battery} \cite{hao2014aggregate, zhao2017geometric}, we show that the building-level controller solves a QP at each time step. The building-level also updates the model parameters with PEM. In this section, all the variables refer to those at the building-level, and thus, we drop the superscript $(i)$ for more concise notation. 

\subsubsection{System Dynamics}
The temperature dynamics of an individual TCL is commonly modeled with Eq. \ref{eq:sys_dynamics}, where $T_t$ is the TCL temperature, $T_{a,t}$ is the ambient temperature, and $q_t\in\{0, 1\}$ is a binary variable representing the operating state, i.e. \textit{on} or \textit{off}, at time t. $P_m$ is the rated power of the TCL. Denoting the thermal resistance and capacitance of the TCL as $R$ and $C$ respectively, the model parameters can be calculated as: $a=\exp\{-\Delta T/(RC)\}$ and $b_t = \eta_t R$, where $\Delta T$ is the time step and $\eta$ is the coefficient of performance (COP). It is challenging to analyze the system dynamics in Eq. \ref{eq:sys_dynamics} due to its nonlinearity. It is common to apply convex relaxation to Eq. \ref{eq:sys_dynamics}, which gives us Eq. \ref{eq:sys_relaxed} \cite{koch2011modeling, hao2014aggregate, zhao2017geometric}. Here $P_t\in[0, P_m]$ is a continuous variable, instead of a binary one.
\begin{subequations}
\begin{equation}\label{eq:sys_dynamics}
    T_{t+1} = a T_t + (1-a)(T_{a,t}-b_t q_t P_m)
\end{equation}
\begin{equation}\label{eq:sys_relaxed}
    T_{t+1} = a T_t + (1-a)(T_{a,t}-b_t P_t)
\end{equation}
\end{subequations}

\subsubsection{Virtual Battery Model}
We abstract the thermal inertia of each TCL with the virtual battery model. Note that our virtual battery model differs from \cite{zhao2017geometric} in that we model the thermal energy instead of electric energy stored in the TCL to account for time-vary COP of the system. With a change of variables $x_t = C(T_{sp}-T_t)$ and $u_t = \eta P_t-Q_{0, t}$, we get Eq. \ref{eq:virtual_battery} from Eq. \ref{eq:sys_relaxed}, where x denotes the state of charge of the virtual battery and $u$ denotes the charging (+) and discharging (-) action. $T_{sp}$ is the setpoint, $\delta = (1-a)RC$, and $Q_{0, t} = (T_{a,t}-T_{sp})/R$ is the nominal thermal flux to keep the TCL temperature at setpoint. The TCL dynamics over a planning horizon is thus characterized by Eq.~\ref{eq:vb_evolve} and can be condensed to AX=BU+C.  
\begin{equation}\label{eq:virtual_battery}
    x_{t+1} = a x_t + u_t \delta
\end{equation}
\begin{equation}\label{eq:vb_evolve}
\resizebox{\hsize}{!}{
$
\underbrace{\begin{bmatrix}
    1  &  &  &  \\
    -a & 1 & & \\
       & \ddots   &\ddots & \\
     &    &   -a  &  1
\end{bmatrix}}_A
\underbrace{\begin{bmatrix}
x_{t+1}\\
x_{t+2}\\
\vdots\\
x_{t+T}
\end{bmatrix}}_X
=\underbrace{\delta \begin{bmatrix}
    1  &  &  &  \\
     & 1 & & \\
       &    &\ddots & \\
     &    &    &  1
\end{bmatrix}}_B
\underbrace{\begin{bmatrix}
u_{t+1}\\
u_{t+2}\\
\vdots\\
u_{t+T}
\end{bmatrix}}_U
+\underbrace{\begin{bmatrix}
x_{t}\\
0\\
\vdots\\
0
\end{bmatrix}}_C
$}
\end{equation}

\subsubsection{Constraints} Each TCL needs to satisfy the function requirement and respects the operational constraints. In this case, we require the TCL temperature to be within the deadband, i.e. $T_t\in[T_{sp}-\Delta,T_{sp}+\Delta]$. At the same time, the system needs to be operating with in its power limits, i.e. $P_t \in [0, P_m]$. Translated to the virtual battery model, $x_t \in [-C\Delta, C\Delta]$ and $u_t\in[-Q_{0,t}, \eta P_m-Q_{0,t}],\; \forall t$. Combining the system dynamics given in Eq. \ref{eq:vb_evolve}, the aforementioned constraints can be written as Eq. \ref{eq:constraints} \cite{zhao2017geometric}, where $\Lambda = A^{-1}$, $\underline{U} = [-Q_{0,t}]$, $\bar{U} = [\eta P_m-Q_{0,t}]$, $\underline{X} = [-C\Delta]$, and $\bar{X} = [C\Delta]$.
\begin{equation} \label{eq:constraints}
\underline{U} \leq U \leq \bar{U}; \quad \underline{X} \leq \Lambda B U + \Lambda C  \leq \bar{X};
\end{equation}

\subsubsection{ Optimization and Learning} The predicted energy consumption at each building is given by Eq. \ref{eq:energy}, where $\langle 1/\eta_t, u_t \rangle$ is the load shifted by the TCLs compared to the baseline load. Note that each building may have more than one TCL. The objective of the building-level controller is to shift $\Delta P_t$ following the aggregator's command and thus the building-level controller solves the problem defined in Eq.~\ref{eq:QP}, which is a QP. We implement the solver with \texttt{CVXPY} \cite{cvxpy}.
\begin{equation} \label{eq:energy}
\hat{P}_{t} = \hat{P}_{t,net}+ \langle 1/\eta_t, u_t \rangle 
\end{equation}
\begin{equation}\label{eq:QP}
\begin{aligned}
\min_{u_{t:t+T-1}} \quad & \sum_{l=0}^{T-1}\|\Delta P_{t+l}-\langle 1/\eta_{t+l},u_{t+l}, \rangle\|^2_2\\
\textrm{s.t.}\quad &  \underline{U} \leq U \leq \bar{U}; \quad \underline{X} \leq \Lambda B U + \Lambda C  \leq \bar{X};
\end{aligned}
\end{equation}

We update the model parameters, $\kappa$, based on new observations from the environment. Instead of optimizing the system-level objectives, we update $\kappa$ by minimizing the prediction error over energy consumption (Eq.~\ref{eq:pem}). We use Adagrad \cite{duchi2011adaptive} to update $\kappa$ every episode (i.e., day) with learning rate of 0.01. 
\begin{equation}\label{eq:pem}
    \mathcal{L}_\kappa = \sum_t (\hat{P}_t - {P}_t)^2
\end{equation}

\section{Results}
We validate our approach in \texttt{CityLearn} environment described in Section~\ref{sec:environment}. We summarize the performance of our predictive model in Section~\ref{sec:res-pred} and our proposed distributed control strategy in Section~\ref{sec:result}. 
\subsection{CityLearn Environment}
\label{sec:environment}
\texttt{CityLearn} \cite{vazquez2019citylearn} is a simulation environment that models thermal storage units in building clusters. Each building is equipped with a chilled water tank supplied by a heat pump. Optionally, a building may also contain a domestic hot water (DHW) tank supplied by a electric water heater, and a photovoltaic (PV) array. The cost function of the environment is defined as the average of net electricity consumption, 1-load factor, ramping, average daily peak demand, and annual peak demand normalized by those of a RBC. The control actions in \texttt{CityLearn} are the charging / discharging of the thermal storage units, with which one can shift the load. Note that control actions as defined by the environment are continuous, which is different from the common assumption for TCLs. Both the simulation and control time-step are 1 hour. The energy consumption of each building consists of heating load, cooling load and non-shiftable appliance load, minus the generation from the solar panel (if applicable). The heating and cooling demand of each building is pre-calculated by CitySim \cite{robinson2009citysim}, a building energy simulator for urban-scale analysis. At the point of writing, \texttt{CityLearn} environment came with one-year of simulation data for four nine-building clusters from four anonymized climate zones.

\subsection{Performance of Predictive Models}\label{sec:res-pred}

\begin{figure}[ht]
    \centering
    \includegraphics[width = 0.9\linewidth]{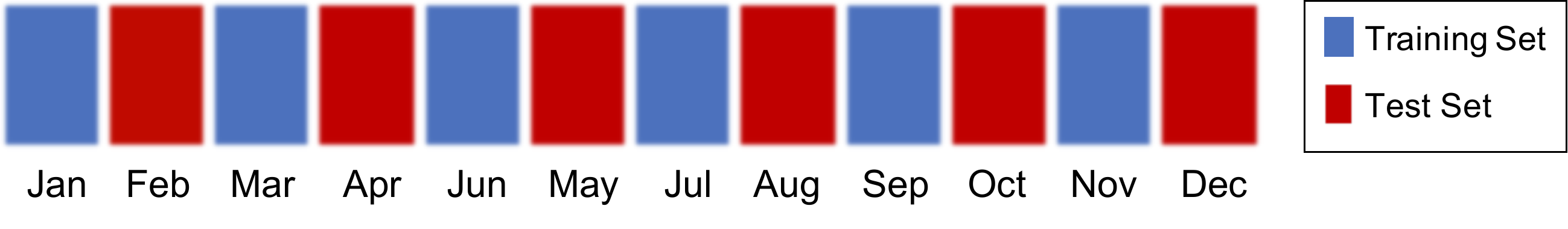}
    \caption{Training (historical) and testing (learning) split.}
    \label{fig:split}
\end{figure}

Even though we evaluate our approach in simulation, we want our experimental set-up to be transferable to real-world. 
We assume historical data are available to pre-train our predictive models. Unfortunately, the current version of \texttt{CityLearn} only provides data over a year and does not provide support for generating additional data. Thus, we split the one-year dataset into training set and test set, as shown in Figure \ref{fig:split}. That is, we split the data between odd and even months, so that the data distribution in the training set and test are similar. We assume the training set to be historical data that is only used to pre-train the predictive models, and  we use the test set to both train and evaluate our proposed learning-based control strategy. The feature used by the predictive model is summarized in Table \ref{tab:feature}.

The output of the predictive model is the total load $P_{total}$, the solar generation $P_{gen}$, and the heating and the cooling load $Q_0$. The performances of predictive models are evaluated by Root-Mean-Squared-Error (RMSE) and Mean-Absolute-Percentage-Error (MAPE) of the predictions for the next 12 hours on test set. The model prediction errors averaged over buildings or climate zones and forecasting horizons are summarized in Table~\ref{tab:pred-results}. The results show that our two predictive models generalize to unseen samples and can trigger accurate load and solar generation predictions over a long horizon.

\begin{table}[h]
\centering
\vspace{0.025in}
\caption{List of features used by the predictive model.}
\label{tab:feature}
\begin{tabular}{ll}\hline 
\textbf{Feature} & \textbf{Description} \\\hline 
\textbf{Building state} & \\
Total load $P_{t, \text{total}}^{(i)}$ & Total electrical load at hour t\\
Indoor temperature(C)  & Indoor temperature \\
Indoor humidity (\%)  & Indoor relative humidity\\
Avg unmet setpoint & Unmet cooling difference \\
Nonshiftable load  (kWh)&  Appliances electricity consumption\\
Solar generation (kWh) & Current solar generation per unit\\
\textbf{Building attribute} &  \\
Building type & Type of building usage \\
Solar power capacity (kW) & Solar power installed\\
DHW demand (KWh)  & Annual domestic hot water demand\\
Cooling demand (kWh)  & Annual cooling demand\\
Electrical demand (kWh)  & Annual electrical demand\\
\textbf{Weather} & \\
Climate zone & Anonymized climate zones \\
Temperature (C)  & Outdoor temperature\\
Outdoor humidity (\%)  & Outdoor relative humidity\\
Diffuse solar radiation & Diffuse solar radiation ($W/m^2$)\\
Direct solar radiation & Direct solar radiation ($W/m^2$)\\
\textbf{Time features} &  \\
Day & Day of year \\
Hour & Hour of day \\
Day type & Type of day from 1 to 8 (holiday) \\
Daylight savings status & Under daylight savings period \\
\hline 
\end{tabular}
\end{table}

\begin{table}[h]
\caption{RMSE and MAPE of predictions on the test set.}
\begin{tabular}{l l l l l}\hline  
 & \textbf{Total load} &  \textbf{Heating} & \textbf{Cooling} & \textbf{Solar}\\ \hline
 \textbf{RMSE} &4.36 \textpm 1.19& 0.07 \textpm 0.04 & 0.04 \textpm 0.01& 47.48 \textpm 1.67\\ 
 \textbf{MAPE} &7.1\% \textpm 2.9\%&12.2\% \textpm 5.7\%&4.2\% \textpm 1.0\%&3.8\% \textpm 0.2\% \\ \hline
\end{tabular}
\label{tab:pred-results}
\end{table}


\subsection{Experiments in CityLearn Environment}\label{sec:result}
We train and evaluate our distributed control solution on the 180-day test set. We repeat the experiment in four nine-building clusters in four climate zones. We initialize $\kappa$ by sampling from a uniform distribution around the ground truth value\footnote{For a model parameter with truth value $\theta$, we initialize $\theta_0\sim \text{Uniform}(0.8\theta, 1.2\theta)$}. We report the cost defined by \texttt{CityLearn} environment of our approach with comparison to other baselines in Table \ref{tab:results}. Each algorithm is evaluated on the test-set for one epoch following the evaluation procedure defined by the \texttt{CityLearn} environment, i.e., executing sequentially on the 180-day test set once. For control strategies with stochasticity, we report the mean and standard deviation of the cost over 5 random seed. Most likely, due to the fact that the cost is evaluated over the entire epoch, the variance is small. The baselines we considered are 1) a no storage scenario, i.e., no load shifting, 2) a RBC controller that charges / discharges the thermal storage based on time defined by the \texttt{CityLearn} environment, 3) a TD3 agent that is provided with the \texttt{CityLearn} environment, and 4) a centralized PPO agent modified from OpenAI gym baselines \cite{baselines}. 

\begin{table}[h!]
\centering
\captionsetup{justification=centering}
\caption{Summary of results. \\ (cost evaluated on the test set for one epoch)}
\begin{tabular}{l c c c c}\hline  
 & \textbf{Climate}&\textbf{Climate}&  \textbf{Climate} & \textbf{Climate}\\& \textbf{1}&\textbf{2}&  \textbf{3} & \textbf{4} \\\cline{2-5}
 &\textbf{(\%)}&\textbf{(\%)}&\textbf{(\%)}&\textbf{(\%)}\\ \hline
 \textbf{No Storage} &100.0&104.4&105.4&104.3\\
\textbf{RBC} &100.0&100.0&100.0&100.0\\ \hline
\textbf{TD3}&104.4 \textpm 0.45& 107.5\textpm 0.62& 110.1\textpm 0.57&108.1 \textpm 0.27\\
\textbf{PPO}&100.7\textpm 0.34&106.5\textpm 0.69 & 105.3\textpm0.71&103.8\textpm0.38\\
\hline
\textbf{Ours} &  \textbf{80.3 \textpm 0.86} & \textbf{83.3 \textpm 3.1} & \textbf{84.5 \textpm 3.1} & \textbf{84.8 \textpm 2.7} \\ \hline
\end{tabular}
\label{tab:results}
\end{table}

\begin{figure}[h]
    \centering
    \includegraphics[width = 1\linewidth]{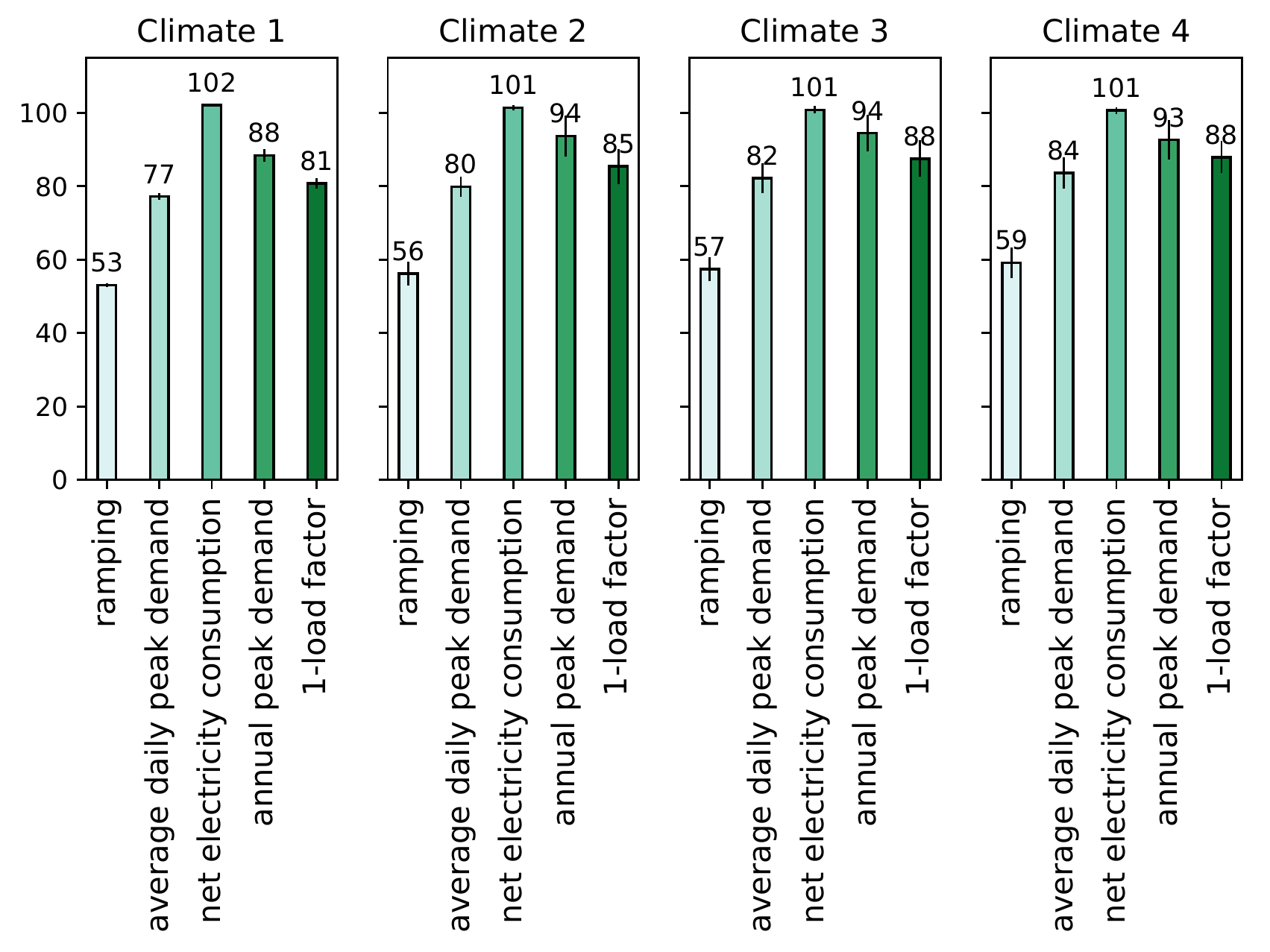}
    \caption{Break down of cost by individual objectives.}
    \label{fig:individual_cost}
\end{figure}

From Table \ref{tab:results}, our approach consistently outperforms all our baselines. On average, we achieved 16.8\% reduction in average cost, compared to the benchmark RBC. It is interesting to note that the two model-free RL baselines do not outperform the RBC in the first epoch. While these model-free RL algorithm outperform our approach asymptotically, with real-world application as the end goal, it is essential that an algorithm does well with limited samples. By incorporating domain knowledge and decomposing the origin problem into more tractable sub-problems, our approach is more sample efficient compared to the model-free RL baselines. 

We also show a breakdown of the overall cost of our approach by individual objectives in Figure \ref{fig:individual_cost}. The pattern of the costs are consistent among four climate zones, indicating that our approach is robust to different climates. Our approach performs particularly well in reducing ramping; average daily peak demand, annual peak demand, and 1-load factor also lowered by 19.3\%, 7.7\%, and 14.6\% respectively. Though net electricity consumption increased by 1.25\%, it is an acceptable compromise for reduced ramping and peak demand.  

\section{Conclusion and Discussion}\label{sec:conclusion}
We've proposed an approach to optimize multiple system-level objectives in the control of a cluster of heterogeneous TCLs and evaluated our approach in a newly-introduced \texttt{CityLearn} environment. We broke down the original problem, which has a large state action space and does not permit an analytical solution, into more tractable sub-problems. We adopt a distributed control approach, which consists of a central load aggregator that optimizes system-level objectives, and building-level controllers that track the target loads planned by the aggregator. We draw inspirations from both RL and MPC to optimize our agents' polices. The aggregator is updated by an ES, a nature-inspired RL algorithm, and the building-level controllers are updated with PEM, a common approach for system identification. We evaluated our approach in four building clusters in four climate zones, and achieved a 16.8\% average reduction in the cost defined by the environment. Our approach also out-performed all four baselines evaluated on the 180-day test set for one epoch. 

In this work, we apportioned the desired load shift to each building with a time-invariant weight vector $\Phi$, following \cite{mahdavi2017model}. However, such approach does not account for the fact that flexibility available at each building could vary with time due to the system characteristics \cite{vrettos2016robust}. It was recently demonstrated in \cite{such2017deep} that ES not only works on shallow neural networks, but also on deep ones. This presents the opportunity to have a more flexibility parameterization of the aggregator's policy. Furthermore, thermal storage units are modeled as perfectly linear with continuous action in the \texttt{CityLearn} environment, which is a over-simplification of realistic systems. We will evaluate our approach in more realistic settings.

\vspace{-0.2 cm}

\bibliographystyle{IEEEbib}
\bibliography{root}

\end{document}